\documentclass[12pt]{article}
\usepackage{graphicx}

      \usepackage[cp1251]{inputenc} 
      \usepackage[russian]{babel}   

\begin{document}
 \noindent {\footnotesize\it Astronomy Letters, 2013, Vol. 39, No. 2, pp. 95--103}

 \noindent
 \begin{tabular}{llllllllllllllllllllllllllllllllllllllllllllll}
 & & & & & & & & & & & & & & & & & & & & & & & & & & & & & & & & & & & & & \\\hline\hline
 \end{tabular}

 \vskip 1.0cm
 \centerline {\large\bf Estimation of the Solar Galactocentric Distance and}
 \centerline {\large\bf Galactic Rotation Velocity from Near-Solar-Circle Objects}
 \bigskip
 \centerline {V.V. Bobylev$^{1,2}$}
 \medskip
{\small\it
 $^1$~Pulkovo Astronomical Observatory, Russian Academy of Sciences

 $^2$~Sobolev Astronomical Institute, St. Petersburg State University, Russia

 }

 \bigskip
{\bf Abstract}---We have tested the method of determining the
solar Galactocentric distance $R_0$ and Galactic rotation velocity
$V_0$ modified by Sofue et al. using near-solar-circle objects.
The motion of objects relative to the local standard of rest has
been properly taken into account. We show that when such young
objects as star-forming regions or Cepheids are analyzed,
allowance for the perturbations produced by the Galactic spiral
density wave improves the statistical significance of the
estimates. The estimate of $R_0=7.25\pm0.32$~kpc has been obtained
from 19 star-forming regions. The following estimates have been
obtained from a sample of 14 Cepheids (with pulsation periods
$P>5^d$): $R_0=7.66\pm0.36$~kpc and $V_0=267\pm17$~km s$^{-1}$. We
consider the influence of the adopted Oort constant $A$ and the
character of stellar proper motions (Hipparcos or UCAC4). The
following estimates have been obtained from a sample of 18
Cepheids with stellar proper motions from the UCAC4 catalog:
$R_0=7.64\pm0.32$~kpc and $V_0=217\pm11$~km s$^{-1}$.

\section{Introduction}

The solar Galactocentric distance $R_0$ and Galactic rotation
velocity $V_0$ are the most important parameters for studying the
structure, kinematics, and dynamics of our Galaxy.

There exist various methods for estimating $R_0$. Reid (1993)
published a review of the $R_0$ determinations made by then by
various methods and obtained the ``best value'' as a weighted mean
of the measurements published over a period of 20 years,
$R_0=8.0\pm0.5$ kpc. Taking into account the main types of errors
and correlations related to the classes of measurements, Nikiforov
(2004) obtained the ``best value'' of $R_0=7.9\pm0.2$ kpc. In the
review by Foster and Cooper (2010), the weighted mean
$R_0=8.0\pm0.4$ kpc was derived by analyzing the $R_0$
determinations published in the last decade. We see that the
present-day value of $R_0$ is known with an error of $0.5\%$.

Individual independent methods can give an estimate of $R_0$ with
an error of $10-15\%$. Note several important individual
measurements. Based on Cepheids and RR Lyr stars belonging to the
bulge (Groenewegen et al. 2008) and using improved calibrations
derived from Hipparcos data and 2MASS photometry, Feast et al.
(2008) found $R_0=7.64\pm0.21$~kpc. By analyzing the orbits of
stars moving around the massive black hole at the Galactic center
(the dynamical parallax method), Gillessen et al. (2009) found
$R_0=8.33\pm0.35$~kpc. According to VLBI measurements, the radio
source Sqr~A* has a proper motion of $6.379\pm0.026$~mas yr$^{-1}$
relative to extragalactic sources (Reid and Brunthaler 2004);
using this value, Sch\"onrich (2012) found $R_0=8.27\pm0.29$~kpc
and $V_0=238\pm9$~km s$^{-1}$. Two H$_{2}$O maser sources, Sgr~B2N
and Sgr~B2M, lie in the immediate vicinity of the Galactic center,
where the source Sgr~A* is located. Based on their direct
trigonometric VLBI measurements, Reid et al. (2009a) obtained
$R_0=7.9^{+0.8}_{-0.7}$~kpc.

The method for estimating $R_0$ and $V_0$ from near-solar-circle
stars is of considerable interest. For example, using only one
maser source, Onsala2, with a high accuracy of determining its
trigonometric parallax, $\approx$3\%, Sofue et al. (2011) obtained
$R_0=7.80\pm0.39$~kpc and $V_0=212\pm10$~km s$^{-1}$. Based on a
sample of seven selected star-forming regions, they found
$R_0=7.54\pm0.77$~kpc. For the star-forming regions, these authors
used rather old kinematic data from Brand and Blitz (1993) without
posing the question about a wide coverage of near-solar-circle
objects.

The goal of this paper is to test the method of determining the
solar Galactocentric distance $R_0$ and Galactic rotation velocity
$V_0$ from young near-solar circle objects. We use the largest
amount of homogeneous data on near-solar-circle objects with the
necessary measurements. These are star-forming regions and young
Cepheids. A peculiarity of our approach is the elimination of the
systematic noncircular stellar motions related to the influence of
the Galactic spiral density wave and a proper allowance for the
motions of objects relative to the local standard of rest.

\section{THE METHOD}

To estimate the Galactocentric distance $R_0$ and the Galaxy’s
circular rotation velocity at the near-solar distance $V_0$ using
near-solar-circle objects, we apply the method proposed by Sofue
et al. (2011):
\begin{equation}
 R_0={r\over 2\cos l}(1-d/r),
\label{R-0}
 \end{equation}
\begin{equation}
 V_0=-{V_p\over 2\cos l}(1-d/r)+V_r\cot l,
\label{V-0}
 \end{equation}
where $r$ is the Galactocentric distance of the star; $d$ is the
distance of the star from the solar circle along the line of sight
(it is desirable that $d\ll r$),
\begin{equation}
 d=-{V_r\over A\sin 2l},
\label{dd-0}
 \end{equation}
$V_r$ is the radial velocity of the star relative to the local
standard of rest; $V_p$ is the velocity perpendicular to $V_r$
directed along the Galactic longitude ($V_p=V_l=4.74r\mu_l\cos b$;
and $A$ is the Oort constant, which is assumed to be
 $A=15$~km s$^{-1}$ kpc$^{-1}$, except for the specially noted cases. Note
that the velocities $V_r$ and $V_p$ should be given relative to
the local standard of rest. The errors of $R_0$ and $V_0$ are
estimated in accordance with the formulas
\begin{equation}
 \delta R_0={1\over 2\cos l}
 \left[\delta r^2+ \left({\delta V_r\over A\sin 2l}\right)^2 \right]^{1/2},
\label{errR-0}
 \end{equation}
\begin{equation}
 \delta V_0={1\over 2\cos l}
 \left[\delta V_p^2+ V_p^2\delta {V_r}^2\left({1\over Ar\sin 2l}-
 {2\cos^2 l\over V_p\sin l}\right)^2 \right]^{1/2}.
\label{errV-0}
 \end{equation}
This method is applicable only for sources located in the first
and fourth Galactic quadrants ($-90^\circ<l<90^\circ$). A
simplified version of Eq.~(1), without the term $d/r,$ is
occasionally used to estimate $R_0$ (Schechter et al. 1992).
According to the approach by Sofue et al. (2011), the fact that a
star belongs to the solar-circle region should be reconciled with
its observed velocities and parameters of the Galactic rotation
curve. Therefore, stringent requirements are placed on the quality
of the observed stellar velocities.

Taking into account the experience of Sofue et al. (2011), we use
the following source selection criteria, having calculated their
Galactocentric distances $R_0$ with a preliminary value of
$R_0=8$~kpc:
\begin{equation}
 -85^\circ<l<85^\circ,
\label{crit-1}
 \end{equation}\begin{equation}
 7~\hbox {kpc}<R<9~\hbox {kpc},
\label{crit-2}
 \end{equation}\begin{equation}
 2.5~\hbox {kpc}<r,
\label{crit-3}
 \end{equation}\begin{equation}
 d/r<1.5
\label{crit-4}
 \end{equation}
without imposing any preliminary constraints on the object’s
radial velocity. Note that the milder criterion~(8) is used for
Cepheids,
 \begin{equation}
 2~\hbox {kpc}<r
 \label{crit-5}
 \end{equation}
to obtain a statistically significant sample. The sample of
Cepheids is of great interest in applying Eqs.~(2) and~(5) using
the stellar proper motions. The number of maser sources with
measured trigonometric parallaxes and proper motions is currently
quite insufficient for this method to be applied.

Equations (1)--(5) were derived by Sofue et al. (2011) under the
assumption of purely circular stellar motions around the Galactic
center. A peculiarity of our approach is the elimination of the
systematic noncircular stellar motions related to the influence of
the Galactic density wave. Such motions are clearly revealed in
the velocities of young objects (Clemens 1985; Bobylev et al.
2008; Bobylev and Bajkova 2010; Stepanishchev and Bobylev 2011).
The following formulas serve as a basis for taking into account
the above effects:
 \begin{equation}
 \begin{array}{rll}
 V_r&=&-u_\odot\cos b\cos l\\
    &-&v_\odot\cos b\sin l-w_\odot\sin b\\
    &+& f_r(GR)\\
    &+&\tilde{v}_\theta\sin(l+\theta)\cos b\\
    &-&\tilde{v}_R \cos(l+\theta)\cos b\\
    &+& V',
 \label{EQ-1}
 \end{array}
 \end{equation}
 \begin{equation}
 \begin{array}{rll}
 V_p&=& u_\odot\sin l-v_\odot\cos l\\
  &+& f_p(GR)\\
  &+&\tilde{v}_\theta \cos(l+\theta)+\tilde{v}_R\sin(l+\theta)\\
  &+& V',
 \label{EQ-2}
 \end{array}
 \end{equation}
where ($u_\odot,v_\odot,w_\odot$) is the group velocity of the
stars under consideration caused by the Sun’s peculiar motion;
$f_r$~(GR) and $f_p$~(GR) denote the functions describing the
differential Galactic rotation, whose specific form is unimportant
our case; and $V'$ denotes the influence of the residual effects.

To take into account the influence of the spiral density wave, we
use the simplest kinematic model based on the linear theory of
density waves by Lin and Shu (1964), where the potential
perturbation has the form of a traveling wave. Then,
 \begin{equation}
 \begin{array}{rll}
      \tilde{v}_R&=&f_R \cos \chi,\\
 \tilde{v}_\theta&=&f_\theta \sin \chi,
 \label{VR-Vtheta}
 \end{array}
 \end{equation}
where $f_R$ and $f_\theta$ are the perturbation amplitudes of the
radial (directed toward the Galactic center in the arm) and
azimuthal (directed along the Galactic rotation) velocities; the
wave radial phase $\chi$ is
 \begin{equation}
   \chi=m[\cot (i)\ln (R/R_0)-\theta]+\chi_\odot,
 \label{chi-creze}
 \end{equation}
$i$ is the spiral pitch angle ($i<0$ for winding spirals); $m$ is
the number of arms, we take $m=2$ here; $\theta$ is the star’s
position angle (measured in the direction of Galactic rotation);
$\chi_\odot$ is the radial phase angle of the Sun measured here
from the center of the Carina-Sagittarius spiral arm
($R\approx7$~kpc). The parameter $\lambda$ is the distance (along
the Galactocentric radial direction) between the adjacent segments
of the spiral arms in the solar neighborhood (spiral wave length)
calculated from the relation $\tan(i)=\lambda m/(2\pi R_0)$.

Initially, we do not constrain the radial velocity, as distinct
from the approach by Sofue et al. (2011), who used the constraint
$|V_r|<15$~km s$^{-1}$. This is because apart from purely circular
motions, the stars have a space velocity dispersion (enters into
the residual velocity $V'$ in Eqs. (11)--(12), which is
$\approx8$~km s$^{-1}$ for star-forming regions and $\approx14$~km
s$^{-1}$ for Cepheids. Therefore, stars with absolute values of
their radial velocities reaching $\approx40$~km s$^{-1}$ contain
useful information. At the final stage, we check the result based
on the $3\sigma$ criterion to eliminate the outliers.

 \begin{figure}[t] {\begin{center}
 \includegraphics[width=110mm]{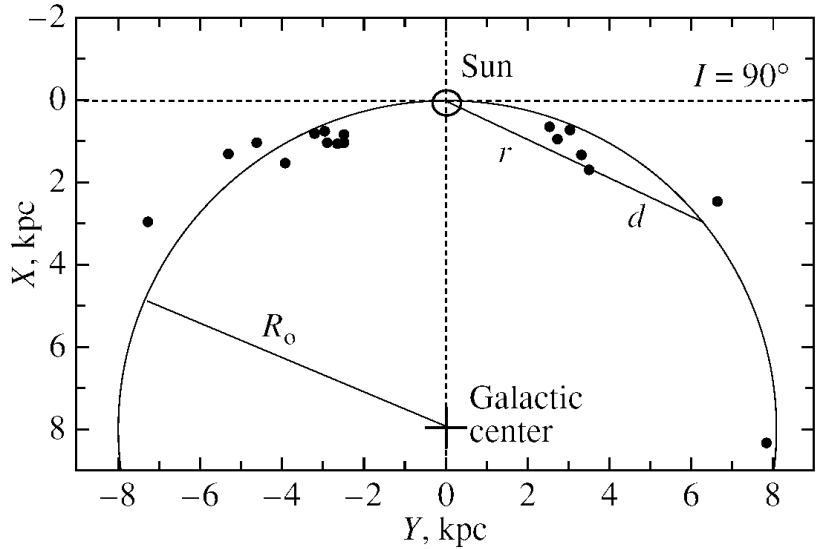}
 \caption{The distribution of star-forming regions in the
Galactic $XY$ plane.}
 \label{HII-19}
 \end{center} }
 \end{figure}

\section{DATA}

We used the radial velocities and photometric distance estimates
for star-forming regions from the catalog by Russeil (2008). All
of the regions satisfying our selection criteria are listed in
Table 1, where alternative names are given in parentheses. These
four regions enter into the sample of seven regions used by Sofue
et al. (2011) to find $R_0= 7.54\pm0.77$~kpc. In addition, the
catalog by Russeil (2008) provides an estimate of the photometric
distance for the region BWW 324, $r=2.1\pm0.6$~kpc; therefore, it
does not enter into our sample. The remaining two regions used by
Sofue et al. (2009), BWW 287 and BWW 328, do not enter into the
catalog by Russeil (2003). The only exception is the region W 49
farthest from the Sun without any distance estimate in the catalog
by Russeil (2003). Mois\'es et al. (2011) estimated the
spectrophotometric distance for this region, $r=12.76\pm4.5$~kpc,
using infrared photometry for a single star. However, following
Sofue et al. (2011), we use a more accurate distance estimate,
$r=11.4\pm1.2$~kpc, obtained by Gwinn et al. (1992) from VLBI
observations of several water masers and the radial velocity
$V_r~(LSR)=5.4\pm2.9$~km s$^{-1}$ determined by Roshi et al.
(2006). The distribution of the selected star-forming regions in
the Galactic plane is presented in Fig. 1, where, as an example,
the distances $r$ and $d$ for one of the regions are shown.

We used data on $\approx$200 classical Cepheids with proper
motions from an improved version of the Hipparcos catalog (ESA,
1997; van Leeuwen 2007). The data from Mishurov et al. (1997),
Gontcharov (2006), and the SIMBAD database served as the sources
of radial velocities. To calculate the distances to Cepheids, the
calibration from Fouqu et al. (2007) is used: $\langle
M_V\rangle=-1.275-2.678\cdot\log P,$ where the period is in days.
Given $\langle M_V\rangle,$ and taking the period-averaged
apparent magnitudes $\langle V\rangle$ and extinction
$A_V=3.23\cdot E(\langle B\rangle-\langle V\rangle)$ mainly from
Acharova et al. (2012) and, for several stars, from Feast and
Whitelock (1997), we determine the distance $r$ from the relation
 \begin{equation}\displaystyle
 r=10^{\displaystyle -0.2(\langle M_V\rangle-\langle V\rangle-5+A_V)},
 \label{Ceph-02}
 \end{equation}
and then assume the relative error in the distances to Cepheids
determined by this method to be 10\%. We used the following
constraint on the absolute value of the residual velocity (after
the subtraction of the Galactic rotation parameters):
$|V_{UVW}|<60$~km s$^{-1}$. Bobylev and Bajkova (2012) used this
same sample of Cepheids to determine the parameters of the
Galactic spiral density wave.

Fourteen Cepheids with proper motions from the Hipparcos catalog
satisfy criteria (6)--(9), given (10); most of them (10 stars) are
long-period ones, with pulsation periods $P\geq9^d,$ i.e., fairly
young stars with a mean age of $\approx55$~Myr (Bobylev and
Bajkova 2012).

\begin{table}[t]                              
\caption[]{\small\baselineskip=1.0ex
  Data on star-forming regions
  }
\begin{center}
      \label{t:88}
\begin{tabular}{|l|rr|c|r|c|r|}\hline
 No.(Russeil, 2003)   & $l^\circ$ & $b^\circ$ & $r\pm\delta r,$~kpc & $V_r$,~km/s & $d/r$  & $R_0\pm \delta R_0,$~kpc  \\\hline
 85~(W49)     & $ 43.2$ & $ 0.0$ & $11.4\pm1.2$ & $ -2.1\pm2.9$ & $ .01$ & $ 7.72\pm1.01$~~ \\
 105          & $ 65.0$ & $ 0.5$ & $ 3.9\pm0.5$ & $ 21.5\pm3.0$ & $ .48$ & $ 6.83\pm0.59$~~ \\
 108          & $ 68.1$ & $ 0.9$ & $ 3.6\pm1.1$ & $  0.6\pm3.0$ & $ .02$ & $ 4.91\pm1.30$~~ \\
 112          & $ 70.3$ & $ 1.6$ & $ 7.0\pm1.5$ & $-30.6\pm3.0$ & $ .46$ & $ 5.61\pm1.84$~~ \\
 114          & $ 71.6$ & $ 2.8$ & $ 2.8\pm0.8$ & $ 11.6\pm3.0$ & $ .46$ & $ 6.48\pm1.04$~~ \\
 116~(S104)   & $ 77.0$ & $ 0.5$ & $ 3.1\pm0.9$ & $ -2.8\pm3.0$ & $ .14$ & $ 5.94\pm1.38$~~ \\
 117          & $ 77.4$ & $-3.7$ & $ 2.6\pm0.8$ & $  6.2\pm3.0$ & $ .37$ & $ 8.19\pm1.25$~~ \\
 310          & $281.8$ & $-2.0$ & $ 4.7\pm1.0$ & $  1.1\pm3.0$ & $ .04$ & $11.05\pm1.60$~~ \\
 314          & $282.6$ & $-1.9$ & $ 3.1\pm0.5$ & $ -0.9\pm3.0$ & $ .04$ & $ 7.41\pm0.82$~~ \\
 320          & $284.0$ & $-1.0$ & $ 3.3\pm0.7$ & $ -3.1\pm3.0$ & $ .13$ & $ 7.72\pm1.05$~~ \\
 321          & $284.2$ & $-0.3$ & $ 5.5\pm0.3$ & $  6.9\pm3.0$ & $ .18$ & $ 9.25\pm0.47$~~ \\
 328          & $287.5$ & $-0.5$ & $ 2.6\pm0.3$ & $-16.8\pm3.0$ & $ .75$ & $ 7.57\pm0.42$~~ \\
 329~(BBW311) & $288.2$ & $-2.9$ & $ 3.1\pm0.9$ & $ -4.4\pm3.0$ & $ .16$ & $ 5.75\pm1.17$~~ \\
 332          & $289.5$ & $ 0.2$ & $ 2.9\pm0.2$ & $-19.9\pm3.0$ & $ .73$ & $ 7.51\pm0.36$~~ \\
 333~(BBW323) & $290.4$ & $-2.9$ & $ 4.2\pm0.5$ & $-12.6\pm3.0$ & $ .31$ & $ 7.87\pm0.65$~~ \\
 334          & $290.4$ & $ 1.5$ & $ 2.7\pm0.3$ & $-15.8\pm3.0$ & $ .60$ & $ 6.18\pm0.39$~~ \\
 336          & $290.6$ & $ 0.2$ & $ 2.8\pm0.3$ & $-21.8\pm3.0$ & $ .79$ & $ 7.12\pm0.39$~~ \\
 338          & $291.3$ & $-0.7$ & $ 2.7\pm0.2$ & $-21.3\pm3.0$ & $ .78$ & $ 6.60\pm0.34$~~ \\
 341          & $291.6$ & $-0.7$ & $ 7.9\pm0.3$ & $ 20.0\pm3.0$ & $ .25$ & $ 8.08\pm0.38$~~ \\
\hline
   Mean, ${\overline R}_0$   &&&&&& $ 7.25$~~  \\
 dispersion, $\sigma_{R_0}$      &&&&&& $ 1.41$~~  \\
    error, $\varepsilon_{R_0}$ &&&&&& $ 0.32$~~  \\\hline
\end{tabular}
\end{center}
 {\small  Note. The radial velocities $V_r$ are given relative to the local
standard of rest (Sch\"onrich et al. 2010) and were corrected for
the influence of the spiral density wave.}
\end{table}

\section{RESULTS AND DISCUSSION}
\subsection{Star-Forming Regions}

Initially, using data on 19 star-forming regions, we obtained a
solution where the radial velocities are given relative to the
standard value for the local standard of rest,
$(U_\odot,V_\odot,W_\odot)=(10.3,15.3,7.7)$~km s$^{-1}$. In this
case, the radial velocities were taken directly from the catalog
by Russeil (2003). The mean value of $R_0$ and dispersion
$\sigma_{R_0}$ are
 \begin{equation}
 R_0= 8.04~\hbox {kpc},~~ \sigma_{R_0}=2.07~\hbox {kpc.}
 \label{HII-R-1}
 \end{equation}
Then, we formed the radial velocities relative to the local
standard of rest with values from Sch\"onrich et al. (2010):
$(U_\odot,V_\odot,W_\odot)=(11.1,12.2,7.3)\pm(0.7,0.5,0.4)$~km
s$^{-1}$. When these velocities were determined, the stellar
metallicity gradient in the Galactic disk and the radial mixing of
stars in the disk were taken into account and modeled. Therefore,
the values of these components currently seem most plausible and
are widely used by various authors. Then,
 \begin{equation}
 R_0= 7.42~\hbox {kpc},~~ \sigma_{R_0}=1.73~\hbox {kpc.}
 \label{HII-R-2}
 \end{equation}
Finally, we corrected the radial velocities derived at the
previous stage for the influence of the spiral density wave using
the following two-armed spiral pattern parameters: the velocity
perturbation amplitudes $f_R=-9$~km s$^{-1}$ and $f_\theta= 0$~km
s$^{-1}$, the pitch angle $i=-5^\circ$ (corresponding to the
wavelength $\lambda=2$~kpc). These parameters are close to those
found by analyzing both maser sources (Bobylev and Bajkova 2010;
Bajkova and Bobylev 2012) and Cepheids (Bobylev and Bajkova 2012).

 \begin{figure}[t] {\begin{center}
 \includegraphics[width=80mm]{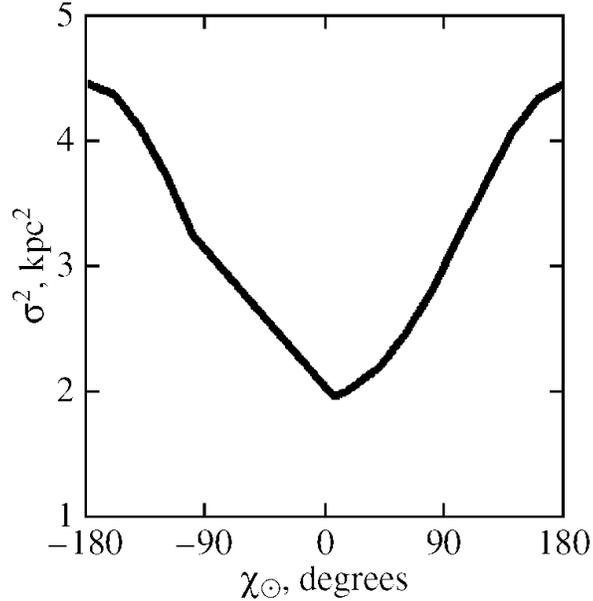}
 \caption{Dispersion squared versus phase of the Sun in the
spiral wave for star-forming regions.}
 \label{HII-chi-2}
 \end{center} }
 \end{figure}

The phase of the Sun in the spiral wave $\chi_\odot$ for
star-forming regions is known poorly. To determine its suitable
value, we analyzed the dispersion squared $\sigma^2$ derived at
all fixed parameters except the phase when solving Eq. (1). The
results are presented in Fig. 2, based on which we adopted the
Sun’s phase $\chi_\odot=-5^\circ$. At this value, $\sigma^2$
reaches its minimum. This results is also of interest in its own
right, because it fits well into the sequence of solar phases as a
function of stellar age (Bobylev and Bajkova, 2012): $-91^\circ,$
$-148^\circ,$ $-193^\circ$ and $-234^\circ$ for stars with ages of
$8,$ $55,$ $95,$ and $135$~Myr. Thus, $\chi_\odot=-5^\circ$ can be
associated with the youngest objects of almost zero age. The fact
that the radial velocities of star-forming regions in the catalog
by Russeil (2003) were determined from the gas component can serve
as grounds for this association.

The results of the last step are presented in Table~1. Column 1
gives the ordinal number from the catalog by Russeil (2003);
columns 2 and 3 contain the Galactic coordinates $l$ and $b$;
column 4 gives the heliocentric distance r; column 5 provides the
radial velocity $V_r$ relative to the local standard of rest
corrected for the influence of the spiral density wave with the
above parameters, the measurement errors of the radial velocities
(except for the region W~49) were taken to be $\delta
V_r=\pm3.0$~km s$^{-1}$; column 6 contains the $d/r$ ratio; and
column 7 gives the $R_0$ estimate. Based on the $3\sigma$
criterion, we rejected region no. 343 (not listed in Table~1).
Comparison of the mean value in the lower row of Table~1 with
results (16) and (17) shows that the dispersion $\sigma_{R_0}$
decreased considerably. The lowest row of the table gives the
error of the mean, $\varepsilon_{R_0}=\sigma_{R_0}/\sqrt{n}$.

\subsection{Cepheids}

Using data on 14 Cepheids, we obtained a solution where the
stellar velocities $V_r$ and $V_p$ are given relative to the local
standard of rest with values from Sch\"onrich et al. (2010). Then,
 \begin{equation}
  \begin{array}{rll}
      R_0&=&7.72~\hbox {kpc},     ~~~~\sigma_{R_0}=1.42~\hbox {kpc,}\\
      V_0&=&255~\hbox {km s$^{-1}$}, ~\sigma_{V_0}=65~\hbox {km s$^{-1}$.}
  \label{Ceph-RV1}
  \end{array}
 \end{equation}
Subsequently, we corrected the stellar velocities $V_r$ and $V_p$
derived at the previous stage for the influence of the spiral
density wave. For this purpose, we used the following two-armed
spiral pattern parameters: the velocity perturbation amplitudes
$f_R=-9$~km s$^{-1}$ and $f_\theta= 0$~km s$^{-1}$, the pitch
angle $i=-5^\circ$ (wavelength $\lambda=2$~kpc), and the phase of
the Sun in the spiral wave $\chi_\odot=-150^\circ$. These are
close to the parameters of the sample of young Cepheids found by
Bobylev and Bajkova (2012).

The results are presented in Table 2. Two stars were rejected
according to the $3\sigma$ criterion: HIP 47177 and HIP 103433
(not listed in Table 2). The velocities $V_r$ and $V_p$ are given
relative to the local standard of rest and were corrected for the
influence of the spiral density wave with the above parameters.
Comparison of the corresponding mean values in the lower row of
Table 2 with solution (18) shows a slight decrease in the errors.

We can see that the dispersion $\sigma_{R_0}$ in our case (Tables
1 and 2) is twice the value of $\sigma_{R_0}=0.77$~kpc derived by
Sofue et al. (2011). Since we imposed no stringent constraints on
the radial velocities of sources, used the largest amount of
available homogeneous data on near-solar-circle objects, and took
into account the systematic effects in the kinematic data, we
obtained a more objective estimate of the method.

\begin{table}[t]                                  
\caption[]{\small\baselineskip=1.0ex
  Parameters found from Cepheids using stellar proper
motions from the Hipparcos catalog
  }
\begin{center}
      \label{t:99}
\begin{tabular}{|c|rr|r|r|r|c|c|c|}\hline
 HIP   & $l^\circ$ & $b^\circ$ & $r,$~ & $V_r$,~~ & $V_p$,~~ & $d/r$ & $R_0\pm \delta R_0,$ & $V_0\pm \delta V_0,$ \\
       &           &           & kpc    &   km/s  &   km/s   &       &                  kpc &                km/s  \\\hline

  96596 & $ 66.9$ & $ 5.3$ & $ 4.2$ & $ 19.0\pm0.3$ & $-112\pm40$ & $ .41$ & $ 7.64\pm0.14$ & $209\pm45$ \\
  98852 & $ 71.1$ & $ 1.4$ & $ 2.4$ & $ 18.5\pm0.3$ & $ -61\pm17$ & $ .83$ & $ 6.85\pm0.15$ & $179\pm21$ \\
  51894 & $284.8$ & $ 2.0$ & $ 3.1$ & $ -2.0\pm1.0$ & $-112\pm14$ & $ .09$ & $ 6.53\pm0.20$ & $239\pm21$ \\
  51262 & $285.6$ & $-1.4$ & $ 2.7$ & $-16.6\pm7.4$ & $ -97\pm 9$ & $ .80$ & $ 8.96\pm0.74$ & $331\pm50$ \\
  50722 & $285.8$ & $-3.3$ & $ 2.8$ & $-15.8\pm5.0$ & $-118\pm11$ & $ .72$ & $ 8.88\pm0.51$ & $379\pm40$ \\
  52570 & $286.0$ & $ 2.4$ & $ 2.2$ & $-12.2\pm4.0$ & $ -97\pm12$ & $ .71$ & $ 6.71\pm0.42$ & $307\pm35$ \\
  53397 & $289.3$ & $-1.2$ & $ 3.9$ & $-27.8\pm5.0$ & $-123\pm15$ & $ .75$ & $10.44\pm0.46$ & $337\pm29$ \\
  54101 & $290.3$ & $-0.8$ & $ 2.5$ & $-14.2\pm4.0$ & $ -92\pm13$ & $ .59$ & $ 5.66\pm0.37$ & $215\pm25$ \\
  54891 & $291.1$ & $ 0.6$ & $ 3.3$ & $-33.2\pm1.0$ & $ -99\pm11$ & $ .99$ & $ 9.21\pm0.16$ & $286\pm14$ \\
  53536 & $291.3$ & $-4.9$ & $ 3.4$ & $-26.2\pm4.0$ & $-111\pm11$ & $ .75$ & $ 8.29\pm0.36$ & $280\pm21$ \\
  53945 & $291.4$ & $-3.9$ & $ 2.7$ & $-21.4\pm5.0$ & $ -96\pm 9$ & $ .77$ & $ 6.60\pm0.43$ & $242\pm24$ \\
  57130 & $294.2$ & $ 2.7$ & $ 5.5$ & $-12.6\pm5.0$ & $-138\pm57$ & $ .21$ & $ 8.05\pm0.40$ & $209\pm64$ \\
  57884 & $296.8$ & $-3.2$ & $ 3.3$ & $-29.8\pm2.7$ & $ -93\pm15$ & $ .74$ & $ 6.44\pm0.23$ & $195\pm18$ \\
  59575 & $298.4$ & $ 0.4$ & $ 3.0$ & $-46.1\pm0.6$ & $-130\pm28$ & $1.22$ & $ 7.01\pm0.13$ & $328\pm28$ \\
  \hline
       Mean &       &  & & &              &        & $ 7.66$ & $267$ \\
 dispersion &       &  & & &              &        & $ 1.35$ & $ 63$ \\
      error &       &  & & &              &        & $ 0.36$ & $ 17$ \\\hline
\end{tabular}
\end{center}
 {\small
Note. The errors of the heliocentric distances $r$ were taken to
be 10\%; the velocities $V_r$ and $V_p$ are given relative to the
local standard of rest (Sch\"onrich et al. 2010) and were
corrected for the influence of the spiral density wave.}
\end{table}

\begin{table}[t]                                  
\caption[]{\small\baselineskip=1.0ex
  Parameters found from Cepheids using stellar
proper motions from the UCAC4 catalog
  }
\begin{center}
      \label{t:909}
\begin{tabular}{|c|rr|r|r|r|c|c|c|}\hline
 star & $l^\circ$ & $b^\circ$ & $r,$~ & $V_r$,~~ & $V_p$,~~ & $d/r$ & $R_0\pm \delta R_0,$ & $V_0\pm \delta V_0,$ \\
        &           &           & kpc    &   km/s  &   km/s   &       &                  kpc &                km/s  \\\hline

  S~Vul & $ 63.4$ & $ 0.8$ & $ 3.6$ & $ 32.0\pm2.0$ & $ -72\pm22$ & $ .73$ & $ 7.05\pm0.19$ & $157\pm24$ \\
 DG~Vul & $ 65.0$ & $-0.9$ & $ 2.1$ & $ 33.3\pm4.0$ & $ -56\pm64$ & $1.39$ & $ 5.89\pm0.33$ & $175\pm71$ \\
  96596 & $ 66.9$ & $ 5.3$ & $ 4.2$ & $ 19.0\pm0.3$ & $-119\pm22$ & $ .41$ & $ 7.64\pm0.14$ & $222\pm25$ \\
  98852 & $ 71.1$ & $ 1.4$ & $ 2.4$ & $ 18.5\pm0.3$ & $ -78\pm12$ & $ .83$ & $ 6.85\pm0.15$ & $228\pm15$ \\
  51894 & $284.8$ & $ 2.0$ & $ 3.1$ & $ -2.0\pm1.0$ & $-127\pm15$ & $ .09$ & $ 6.53\pm0.20$ & $271\pm23$ \\
 CS~Car & $285.6$ & $ 0.2$ & $ 3.4$ & $ -6.1\pm3.0$ & $ -70\pm 5$ & $ .23$ & $ 7.86\pm0.33$ & $162\pm13$ \\
  51262 & $285.6$ & $-1.4$ & $ 2.7$ & $-16.6\pm7.4$ & $ -97\pm11$ & $ .80$ & $ 8.96\pm0.74$ & $331\pm51$ \\
  50722 & $285.8$ & $-3.3$ & $ 2.8$ & $-15.8\pm5.0$ & $ -63\pm17$ & $ .72$ & $ 8.88\pm0.51$ & $204\pm30$ \\
  52570 & $286.0$ & $ 2.4$ & $ 2.2$ & $-12.2\pm4.0$ & $ -82\pm12$ & $ .71$ & $ 6.71\pm0.42$ & $261\pm31$ \\
 II~Car & $288.2$ & $-0.7$ & $ 6.8$ & $  7.4\pm3.0$ & $-141\pm46$ & $ .12$ & $ 9.52\pm0.31$ & $196\pm60$ \\
  53397 & $289.3$ & $-1.2$ & $ 3.9$ & $-27.8\pm5.0$ & $ -88\pm10$ & $ .75$ & $10.44\pm0.46$ & $244\pm20$ \\
  54101 & $290.3$ & $-0.8$ & $ 2.5$ & $-14.2\pm4.0$ & $-107\pm16$ & $ .59$ & $ 5.66\pm0.37$ & $250\pm29$ \\
  54891 & $291.1$ & $ 0.6$ & $ 3.3$ & $-33.2\pm1.0$ & $ -83\pm47$ & $ .99$ & $ 9.21\pm0.16$ & $244\pm56$ \\
  53536 & $291.3$ & $-4.9$ & $ 3.4$ & $-26.2\pm4.0$ & $ -82\pm11$ & $ .75$ & $ 8.29\pm0.36$ & $208\pm17$ \\
  53945 & $291.4$ & $-3.9$ & $ 2.7$ & $-21.4\pm5.0$ & $ -64\pm 9$ & $ .77$ & $ 6.60\pm0.43$ & $165\pm19$ \\
  57130 & $294.2$ & $ 2.7$ & $ 5.5$ & $-12.6\pm5.0$ & $-138\pm85$ & $ .21$ & $ 8.05\pm0.40$ & $208\pm95$ \\
  57884 & $296.8$ & $-3.2$ & $ 3.3$ & $-29.8\pm2.7$ & $ -93\pm30$ & $ .74$ & $ 6.44\pm0.23$ & $195\pm32$ \\
  59575 & $298.4$ & $ 0.4$ & $ 3.0$ & $-46.1\pm0.6$ & $ -61\pm47$ & $1.22$ & $ 7.01\pm0.13$ & $168\pm49$ \\
  \hline
       Mean &       &  & & &              &        & $ 7.64$ & $217$ \\
 dispersion &       &  & & &              &        & $ 1.34$ & $ 46$ \\
      error &       &  & & &              &        & $ 0.32$ & $ 11$ \\\hline
\end{tabular}
\end{center}
 {\small
Note. The proper motions for two stars, HIP~51262 and HIP~57884,
remained the same, these were taken from the Hipparcos catalog;
the velocities $V_r$ and $V_p$ are given relative to the local
standard of rest (Sch\"onrich et al. 2010) and were corrected for
the influence of the spiral density wave.}
\end{table}

As can be seen from Table 2, the radial velocities have a better
accuracy than the $V_p$ components in random terms. The errors in
the proper motions affect significantly the accuracy of
determining the $V_p$ component (see Eq. (5) and Table 2). At
present, a new version of the UCAC4 catalog has been published
(Zacharias et al. 2012), where about 140 catalogs were used to
derive the mean stellar proper motions. Therefore, the velocities
$V_0$ calculated using these proper motions are of considerable
interest. We repeated our calculations for the sample of 14
Cepheids for the case where their proper motions were taken from
the UCAC4 catalog. It turned out to be necessary to retain the
proper motions from the Hipparcos catalog for two stars, HIP~51262
and HIP~57884. Based on this combined sample of Cepheids (with the
UCAC4 proper motions for 12 stars and the Hipparcos proper motions
for two stars), we obtained the following parameters:
 \begin{equation}
  \begin{array}{rll}
      R_0&=&7.66~\hbox {kpc}, ~~~~~\sigma_{R_0}=1.35~\hbox {kpc,}\\
      V_0&=&229~\hbox {km s$^{-1}$}, ~\sigma_{V_0}=44~\hbox {km s$^{-1}$,~~for}\\
        A&=& 15~\hbox {km s$^{-1}$ kpc$^{-1}$,}
  \label{Ceph-A15}
  \end{array}
 \end{equation}
They have slightly smaller random errors than those in Table 2. We
also see that $V_0$ decreased by $\approx1\sigma$. Based on the
same sample of Cepheids, we repeated solution (19) for two
different values of the Oort constant $A$:
 \begin{equation}
  \begin{array}{rll}
      R_0&=&7.47~\hbox {kpc}, ~~~~\sigma_{R_0}=1.31~\hbox {kpc,}\\
      V_0&=&224~\hbox {km s$^{-1}$}, ~\sigma_{V_0}=43~\hbox {km s$^{-1}$,~~for}\\
        A&=& 16~\hbox {km s$^{-1}$ kpc$^{-1}$,}
  \label{Ceph-A16}
  \end{array}
 \end{equation}
and
 \begin{equation}
  \begin{array}{rll}
      R_0&=&7.31~\hbox {kpc}, ~~~~\sigma_{R_0}=1.27~\hbox {kpc,}\\
      V_0&=&219~\hbox {km s$^{-1}$}, ~\sigma_{V_0}=42~\hbox {km s$^{-1}$,~~for}\\
        A&=& 17~\hbox {km s$^{-1}$ kpc$^{-1}$.}
  \label{Ceph-A17}
  \end{array}
 \end{equation}
It can be seen that $R_0$ and $V_0$ decrease with increasing Oort
constant $A,$ but their dispersions change insignificantly. The
present-day values of the Oort constant $A$ lie within the range
14--18~km s$^{-1}$ kpc$^{-1}$. For example,
 $A=14.8\pm0.8$~km s$^{-1}$ kpc$^{-1}$ was found using Cepheids from the Hipparcos
catalog (Feast and Whitelock 1997), while its value derived from
44 maser sources with measured trigonometric parallaxes is
$A=16.7\pm0.6$ km s$^{-1}$ kpc$^{-1}$ (Bajkova and Bobylev 2012).

Table 3 presents the results of our analysis for 18 stars with
proper motions from the UCAC4 catalog (except for HIP 51262 and
HIP 57884). It turned out that four more stars, S~Vul, DG~Vul,
CS~Car, and II~Car, have fairly reliable measurements of their
radial velocities and proper motions. Therefore, they were
included in the sample. To find such stars, we identified all
Cepheids ($\approx$450 stars) from the catalog by Berdnikov et al.
(2000), whose distances were determined from infrared photometry,
with the UCAC4 catalog. In particular, the distances to the
Cepheids CS~Car and II~Car were taken from the catalog by
Berdnikov et al. (2000). Not all of the near-solar-circle Cepheids
entered into the sample, because their proper motions are not
always reliable; for example, we encountered the cases where $V_0$
was found to be negative, etc. Just as in other tables, the
calculations were made with the Oort constant $A=15$~km s$^{-1}$
kpc$^{-1}$. On the whole, we can conclude that adding the stellar
proper motions from the UCAC4 catalog had a favorable effect on
the determination of the sought-for parameters, in particular, the
dispersion $\sigma_{V_0}$ decreased.

The values of $R_0$ found (Tables 1, 2, and 3) agree, within
$1\sigma,$ with those discussed in the Introduction.

Our Galactic rotation velocities (Tables 2 and 3) agree, within
$1\sigma,$ with the results of a kinematic analysis for various
samples of stars. For example, the velocity derived from maser
sources with measured trigonometric parallaxes is
 $V_0=254\pm16$~km s$^{-1}$ for $R_0=8.4$ kpc (Reid et al. 2009b),
 $V_0=244\pm13$~km s$^{-1}$ for $R_0=8.2$ kpc (Bovy et al. 2009), or
 $V_0=248\pm14$~km s$^{-1}$ for $R_0=8$ kpc (Bobylev and Bajkova 2010). Analysis of
blue supergiants (Zabolotskikh et al. 2002) and OB associations
(Mel’nik and Dambis 2009) also shows that the velocity
 $V_0=240-250$~km s$^{-1}$ for $R_0=8$ kpc. At the same time, young Cepheids
rotate somewhat more slowly, with
 $V_0=209\pm16$~km s$^{-1}$, than slightly older Cepheids with
 $V_0=243\pm18$~km s$^{-1}$ for $R_0=8$~kpc (Bobylev and Bajkova, 2012).

\section{CONCLUSIONS}

We tested the method of determining the solar Galactocentric
distance $R_0$ and Galactic rotation velocity $V_0$ modified by
Sofue et al. (2011) using data on star-forming regions and young
Cepheids near the solar circle. A peculiarity of this method is
that the belonging of a star to the solar-circle region is
reconciled with its observed velocities and parameters of the
Galactic rotation curve.

We attempted to encompass the largest amount of homogeneous data
on near-solar-circle objects with the necessary measurements. We
properly took into account the motions of objects relative to the
local standard of rest. For this purpose, we used the parameters
of not the standard solar motion relative to the local standard of
rest, which are commonly used by radio astronomers to retain the
continuity of their results, but more reliable, present-day
parameters of this motion. We showed that when young objects are
analyzed, the perturbations produced by the Galactic spiral
density wave should be taken into account, which causes the
dispersion of the estimated parameters to decrease.

We obtained the estimate $R_0=7.25\pm0.32$~kpc from 19
star-forming regions. Based on a sample of Cepheids, we studied
the influence of the adopted Oort constant $A$ and the character
of the stellar proper motions used on the parameters being
determined. For this purpose, we invoked the proper motions from
the Hipparcos and UCAC4 catalogs. Based on a sample of 14
long-period Cepheids with Hipparcos proper motions, we obtained
the following estimates: $R_0=7.66\pm0.36$~kpc and
$V_0=267\pm17$~km s$^{-1}$. Based on a sample of 18 Cepheids
obtained by invoking the UCAC4 stellar proper motions, we found
$R_0=7.64\pm0.32$~kpc and $V_0=217\pm11$~km s$^{-1}$. Note that we
considered almost all of the known Galactic Cepheids with the
necessary photometric and kinematic data.

This independent method may turn out to be useful in the case of a
considerable increase in the amount of kinematically homogeneous
data with highly accurate distance estimates (the distance errors
make a crucial contribution to the dispersion of the results) and
proper motions. These include, for example, Galactic maser sources
whose trigonometric parallaxes are determined by various research
groups by the VLBI method. The number of open star clusters with
known kinematic parameters increases. Finally, as a result of the
accomplishment of the GAIA space mission, the number of distant
stars with measured parallaxes, radial velocities, and proper
motions increases considerably.

\bigskip{\bf ACKNOWLEDGMENTS}\bigskip

I am grateful to the referee for valuable remarks that contributed
to a significant improvement of this paper. This work was
supported in part by the ``Nonstationary Phenomena in Objects of
the Universe'' Program of the Presidium of the Russian Academy of
Sciences and grant NSh--1625.2012.2 from the President of the
Russian Federation. In our study, we used the SIMBAD search
database.

\bigskip{\bf REFERENCES}\bigskip

{\small
 1. I.A. Acharova, Yu. N. Mishurov, and V.V. Kovtyukh, Mon.
Not. R. Astron. Soc. 420, 1590 (2012).

2. A.T. Bajkova and V.V. Bobylev, Astron. Lett. 38, 549 (2012).

3. L.N. Berdnikov, A.K. Dambis, and O.V. Vozyakova, Astron.
Astrophys. Suppl. Ser. 143, 211 (2000).

4. V.V. Bobylev and A.T. Bajkova, Mon. Not. R. Astron. Soc. 408,
1788 (2010).

5. V.V. Bobylev and A.T. Bajkova, Astron. Lett. 38, 638 (2012).

6. V.V. Bobylev, A.T. Bajkova, and A.S. Stepanishchev, Astron.
Lett. 34, 515 (2008).

7. J. Bovy, D.W. Hogg, and H.-W. Rix, Astrophys. J. 704, 1704
(2009).

8. J. Brand, and L. Blitz, Astron. Astrophys. 275, 67 (1993).

9. D.P. Clemens, Astrophys. J. 295, 422 (1985).

10. M. Feast and P. Whitelock, Mon. Not. R. Astron. Soc. 291, 683
(1997).

11. M.W. Feast, C.D. Laney, T.D. Kinman, et al., Mon. Not. R.
Astron. Soc. 386, 2115 (2008).

12. T. Foster and B. Cooper, ASP Conf. Ser. 438, (2010).

13. P. Fouqu, P. Arriagada, J. Storm, et al., Astron. Astrophys.
476, 73 (2007).

14. S. Gillessen, F. Eisenhauer, S. Trippe, et al., Astrophys. J.
692, 1075 (2009).

15. G.A. Gontcharov, Astron. Lett. 32, 759 (2006).

16. M.A.T. Groenewegen, A. Udalski, and G. Bono, Astron.
Astrophys. 481, 441 (2008).

17. C.R. Gwinn, J.M. Moran, and M.J. Reid, Astrophys. J. 393, 149
(1992).

18. The HIPPARCOS and Tycho Catalogues, ESA SP1200 (1997).

19. F. van Leeuwen, Astron. Astrophys. 474, 653 (2007).

20. C.C. Lin and F.H. Shu, Astrophys. J. 140, 646 (1964).

21. A.M. Mel'nik and A.K. Dambis, Mon. Not. R. Astron. Soc. 400,
518 (2009).

22. Yu.N. Mishurov, I.A. Zenina, A.K. Dambis, et al., Astron.
Astrophys. 323, 775 (1997).

23. A.P. Mois\'es, A. Damineli, E. Figuer\'edo, et al., Mon. Not.
R. Astron. Soc. 411, 705 (2011).

24. I.I. Nikiforov, ASP Conf. Ser. 316, 199 (2004).

25. M.J. Reid, Annu. Rev. Astron. Astrophys. 31, 345 (1993).

26. M. Reid and A. Brunthaler, Astrophys. J. 616, 872 (2004).

27. M. Reid, K.M. Menten, X.W. Zheng, et al., Astrophys. J. 705,
1548 (2009a).

28. M. Reid, K.M. Menten, X.W. Zheng, et al., Astrophys. J. 700,
137 (2009b).

29. D.A. Roshi, C.G. De Pree, W.M. Goss, et al., Astrophys. J.
644, 279 (2006).

30. D. Russeil, Astron. Astrophys. 397, 133 (2003).

31. P.L. Schechter, I.M. Avruch, J.A.R. Caldwell, et al., Astron.
J. 104, 1930 (1992).

32. R. Sch\"onrich, arXiv: 1207.3079 (2012).

33. R. Sch\"onrich, J. Binney, and W. Dehnen, Mon. Not. R. Astron.
Soc. 403, 1829 (2010).

34. Y. Sofue, T. Nagayama, M. Matsui, et al., Publ. Astron. Soc.
Jpn. 63, 867 (2011).

35. A.S. Stepanishchev and V.V. Bobylev, Astron. Lett. 37, 254
(2011).

36. M.V. Zabolotskikh, A.S. Rastorguev, and A.K. Dambis, Astron.
Lett. 28, 454 (2002).

37. N. Zacharias, C.T. Finch, T.M. Girard, et al., I/322
Catalogue, Strasbourg Data Base (2012).

}

\end{document}